\documentclass[aps,prb,11pt,a4paper,reprint,amsmath,amssymb,superscriptaddress]{revtex4-2} 
\usepackage{bm}
\usepackage{graphicx}
\usepackage{xcolor}
\usepackage{transparent}
\usepackage{times}
\usepackage{mathptmx}

\newcommand{\wh}{\widehat}
\newcommand{\wt}{\widetilde}

\newcommand{\ave}[1]{{\left<#1\right>}}

\newcommand{\TDE}{{TDE}}

\newcommand{\tauw}{\ensuremath{\tau_\text{w}}}

\newcommand{\tlag}{\triangle_t}
\newcommand{\xlag}{\triangle_x}
\newcommand{\ylag}{\triangle_y}

\newcommand{\aave}{\ensuremath{\langle{a}\rangle}}

\newcommand{\Kave}{\ensuremath{\langle{K}\rangle}}
\newcommand{\Phiave}{\ensuremath{\langle{\Phi}\rangle}}
\newcommand{\Phirms}{\ensuremath{\Phi}_\text{rms}}

\newcommand{\Eqref}[1]{Eq.~\eqref{#1}}
\newcommand{\Eqsref}[1]{Eqs.~\eqref{#1}}
\newcommand{\Figref}[1]{Fig.~\ref{#1}}

\newcommand{\Secref}[1]{Sec.~\ref{#1}}

\begin{document}
\title{A three-point velocity estimation method for turbulent flows in two spatial dimensions}
%{Improved TDE method for turbulence velocity inference}
%{Two-dimensional velocity estimation of uncorrelated structures using coarse-grained imaging diagnostics}

\author{J.~M.~Losada}
\email{juan.m.losada@uit.no}
\affiliation{Department of Physics and Technology, UiT The Arctic University of Norway, N-9037 Troms{\o}, Norway}

\author{A.~D.~Helgeland}
\email{aurora.helgeland@uit.no}
\affiliation{Department of Physics and Technology, UiT The Arctic University of Norway, N-9037 Troms{\o}, Norway}

\author{J.~L.~Terry}
\email{terry@psfc.mit.edu}
\affiliation{MIT Plasma Science and Fusion Center, Cambridge, MA 02139, United States of America}

\author{O.~E.~Garcia}
\email{odd.erik.garcia@uit.no}
\affiliation{Department of Physics and Technology, UiT The Arctic University of Norway, N-9037 Troms{\o}, Norway}

\date{\today}

\begin{abstract}
Time delay and velocity estimation has been a widely studied subject in the context of signal processing, with applications in many different fields of physics. The velocity of fluctuation structures is typically estimated as the distance between two measurement points divided by the time lag that maximizes the cross-correlation function between the measured signals. In this contribution, we demonstrate that this technique is not suitable in two spatial dimensions unless the velocity is aligned with the separation between the measurement points. We present an improved method to accurately estimate both components of the velocity vector relying on three non-aligned measurement points. The cross-correlation based three-point time delay method is shown to give exact results for the velocity components in the case of a super-position of uncorrelated Gaussian pulses. The new technique is tested on synthetic data generated from realizations of such processes for which the underlying velocity components are known. The results are compared with and found vastly superior to those obtained using the standard two-point technique. Finally, we demonstrate the applicability of the three-point method on gas puff imaging data of strongly intermittent plasma fluctuations at the boundary of the Alcator C-Mod tokamak.
\end{abstract}

\maketitle

\clearpage

\section{Introduction}

Velocity estimation plays a crucial role in a wide range of scientific and technological fields. Its applications extend from analyzing turbulent systems like atmospheric phenomena and magnetized plasmas, to clinical ultrasound scanning \cite{kortbek_estimation_2006}, and to exploring disciplines such as astrophysics and space sciences \cite{coles_interplanetary_1978}. Moreover, it plays a key role in enhancing the effectiveness of radar systems and improving communication technologies \cite{kintner_ionosphere_2005}. Many different techniques have been studied conditioned by the field of application and the experimental setup. In this contribution, we are addressing the problem of estimating the velocity of localized structures in an intensity field recorded in a two-dimensional plane.

One particular example of such a system are plasma fluctuations at the boundary of magnetically confined plasmas. In many experimental setups, a gas-puff imaging diagnostic measures light emission from coherent structures that propagate radially out of the plasma column \cite{stotler_neutral_2003,zweben_edge_2007,zweben_invited_2017}. Different approaches have been employed for velocity estimation in this setup. Most commonly the velocity is inferred from time-delay estimation (\TDE) from the measured signals. \TDE\ methods range from cross-correlation techniques \cite{jakubowski_observation_2002,mckee_experimental_2003, holland_investigation_2004,balazs_cross_2011,zoletnik_methods_2012,zweben_comparison_2013,cziegler_fluctuating_2013, cziegler_zonal_2014,prisiazhniuk_magnetic_2017,lampert_novel_2021}, to wavelet methods \cite{jakubowski_wavelet_2001} and dynamic programming 
\cite{gupta_dynamic_2010,shao_velocimetry_2013}. Alternative approaches to velocity estimation, not relying on cross-correlations, encompass Fourier analysis \cite{cziegler_experimental_2010, xu_fourier_2010,cziegler_ion-cyclotron_2012,sierchio_comparison_2016}, optical flow continuity \cite{munsat_derivation_2006,sechrest_flow_2011}, blob tracking \cite{kube_blob_2013,fuchert_blob_2014,zweben_blob_2016,decristoforo_blob_2020} and machine learning based algorithms \cite{han_tracking_2022}.

In its simplest form, the standard method estimates the velocity of propagation between two measurement points based on the distance between those points and the time delay between the signals recorded. In fluctuating media, this time delay is commonly estimated as the time lag that maximizes the cross-correlation function between the signals. However, this method is inaccurate if the velocity of propagation is not aligned with the measurement points or when the propagating structures are tilted \cite{fedorczak_physical_2012}. Fourier analysis methods are subject to the same limitations as this two-point estimation method \cite{sierchio_comparison_2016}. % Others: what about all those collected in the TDE folder on Box Sync/SOL/?
Improved methods for velocity estimation have been developed but are limited to high image resolution \cite{terry_velocity_2005, zweben_estimate_2011, zweben_edge_2015, lampert_novel_2021}. % Others? Agostini, Zweben, etc?

In the following section, we present an improved three-point method to estimate the velocity of fluctuation structures moving in a two-dimensional plane from simple geometric considerations. In \Secref{sec.model} a stochastic model describing the fluctuations as a super-position of Gaussian pulses is introduced. This demonstrates that the three-point method can be used with the time delays estimated from cross-correlation functions, and is independent of the distribution of pulse amplitudes as well as their rate of occurrence. This is verified by analysis of synthetic data from realizations of this process in \Secref{sec.num}. In \Secref{sec.exp} the three-point method applied to measurement data from gas puff imaging experiments at the boundary of the Alcator C-Mod tokamak. A discussion of the new method and conclusions of these investigations are presented in \Secref{sec.dc}. Finally, a Python implementation of the techniques described in this paper is available publicly on GitHub \cite{losada_velocity_estimation}.

\section{Time delay estimation}

\begin{figure}[t]
\centering
\def\svgwidth{0.9\columnwidth}
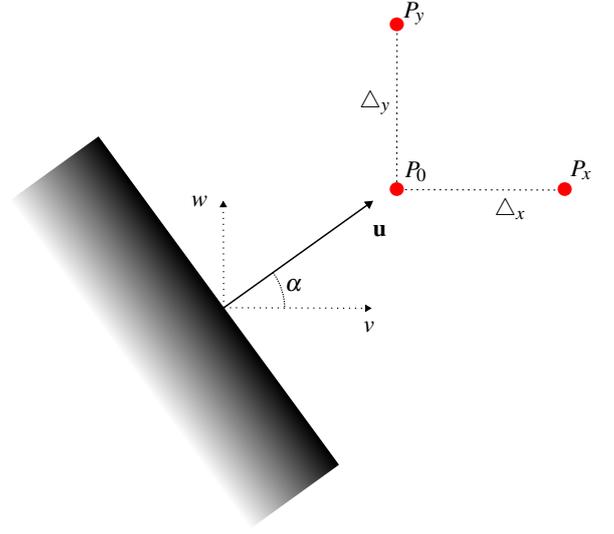
\caption{A structure moving with velocity components $(v,w)$ is recorded at fixed measurement points $P_0$, $P_x$ and $P_y$. $P_x$ is separated a horizontal distance $\xlag$ from $P_0$ and $P_y$ is separated a vertical distance $\ylag$ from $P_0$.}
\label{fig.wave_front}
\end{figure}

Consider a front propagating in a two-dimensional plane with horizontal velocity component $v$ along the $x$-axis and vertical velocity component $w$ along the $y$-axis. The perturbation is measured at three spatially separated points $P_0$, $P_x$ and $P_y$ as illustrated in \Figref{fig.wave_front}. In this arrangement, $P_x$ is horizontally separated from $P_0$ by a distance $\xlag$, while $P_y$ is vertically separated from $P_0$ by a distance $\ylag$. We denote the velocity vector angle from the horizontal axis by $\alpha$.

The objective is to estimate the velocity components $v$ and $w$ from the signals measured at $P_0$, $P_x$ and $P_y$. Consider first the recordings at $P_0$ and $P_x$. Let ${\tau}_x$ be the time delay between the fronts measured at these positions. In the standard two-point \TDE\ method, this time lag is interpreted as the time taken for the pulse to traverse horizontally the distance $\xlag$ between points $P_0$ and $P_x$, that is, ${\tau}_x=\xlag/v$. From this, the two-point estimate for the horizontal velocity component $v$ is given by
\begin{subequations}\label{eq.vw2.hat}
\begin{equation}
    \wh{v}_2 = \frac{\xlag}{\tau_x} , 
\end{equation}
where we used the subscript $2$ to denote the two-point estimate and added a hat symbol $\wh{\cdot}$ to indicate that these relations are meant as estimates of the velocity components. Similarly, by estimating the time lag ${\tau}_y$ between the signals measured at $P_0$ and $P_y$, the vertical velocity component is estimated as
\begin{equation}
    \wh{w}_2 = \frac{\ylag}{\tau_y} .
\end{equation}
\end{subequations}
This is the standard two-point velocity estimation method, which is essentially a one-dimensional treatment of the problem.

However, this interpretation of the time lags ${\tau}_x$ and $\tau_y$ is obviously flawed. To illustrate this, consider the special case where the front moves strictly vertical, rendering $v=0$. Ignoring tilt effects, such a pulse passes through the measurement points $P_0$ and $P_x$ simultaneously, resulting in a time lag ${\tau}_x=0$. This would then lead to an infinitely large estimated horizontal velocity component $\wh{v}_2$. More generally, if the pulse moves at a slight angle to any coordinate axis, the two-point method will give a severe overestimate of the corresponding velocity component.

With three non-aligned measurement points $P_0$, $P_x$ and $P_y$, it is straight forward to give correct estimates of the two velocity components. As the pulse moves, it will first be measured at $P_0$ and some time later at $P_x$ as seen in \Figref{fig.wave_front}. The structure will have travelled a distance $\xlag\cos{\alpha}=v\xlag/(v^2+w^2)^{1/2}$ along its direction of motion between its arrival at $P_0$ and $P_x$. The time lag $\tau_x$ corresponds then to the time $\xlag\cos\alpha/u$ it takes for the pulse travel this distance,
\begin{subequations}\label{eq.tau2}
\begin{equation}\label{eq.taux}
    \tau_x = \frac{v\xlag}{v^2+w^2} .
\end{equation}
In order to estimate both velocity components, a third measurement point which is vertically separated from $P_0$ is required. A similar geometrical consideration gives the distance the structure travels as $\ylag\sin{\alpha}=w\ylag/(v^2+w^2)^{1/2}$, where $\ylag$ is the vertical separation of the measurement points. The resulting lag time is then given by $\ylag\sin\alpha/u$ or in terms of the velocity components
\begin{equation}\label{eq.tauy}
    {\tau}_y = \frac{w\ylag}{v^2+w^2} .
\end{equation}
\end{subequations}
The relations given by \Eqsref{eq.taux} and \eqref{eq.tauy} can be inverted with respect to $v$ and $w$, leading to the velocity component estimates
\begin{subequations}\label{eq.vhat}
\begin{align} 
    \wh{v}_3 &= \frac{{\tau_x}\xlag}{{\tau_x}^2 + {\tau_y}^2 \xlag^2/\ylag^2 } \label{eq.est.v} ,
    \\ 
    \wh{w}_3 &= \frac{{\tau_y}\ylag}{{\tau_y}^2 + {\tau_x}^2 \ylag^2/\xlag^2} . \label{eq.est.w}
\end{align}
\end{subequations}
These estimates do not have the deficiency of the simple two-point method when the pulse moves along one of the coordinate axes. In particular, if the pulse arrives simultaneously at $P_0$ and $P_x$, the time lag $\tau_x$ vanishes and so does the estimated horizontal velocity component from \Eqref{eq.est.v}.

Note that by dividing the numerator and denominator of \Eqref{eq.est.v} by $\xlag^2$ and of \Eqref{eq.est.w} by $\ylag^2$, it is possible to write these estimates in terms of the simple two-point velocity estimates,
\begin{subequations}\label{vw3.hat}
\begin{align}
    \wh{v}_3 &= \frac{\wh{v}_2}{1+(\wh{v}_2/\wh{w}_2)^2} ,
    \\
    \wh{w}_3 &= \frac{\wh{w}_2}{1+(\wh{w}_2/\wh{v}_2)^2} .
\end{align}
\end{subequations}
From these relations it follows that the two-point estimates are always greater than or equal to the three-point estimates, $\wh{v}_2\geq\wh{v}_3$ and $\wh{w}_2\geq\wh{w}_3$. Specifically, when the horizontal and vertical velocity components are equal, the simple two-point method overestimates the velocities by a factor two. Moreover, the ratio of the two velocity components are reversed for the two-point method since $\wh{v}_3/\wh{w}_3=\wh{w}_2/\wh{v}_2$. In particular, if $\wh{v}_3>\wh{w}_3$ then $\wh{v}_2<\wh{w}_2$ and vice versa.

These considerations demonstrate that the two velocity components of a sharp front can be accurately estimated by using three non-aligned measurement points. It is straight forward to generalize this to a localized pulse which is symmetric with respect to its direction of motion, which gives the same expressions for the velocity components. However, in this case, the distance between the measurement points must be smaller than the structure size. Moreover, in turbulent flows, there can be overlap of structures and a distribution of amplitudes, sizes and velocities. This motivates a statistical treatment, which is presented in the following section.

\section{Stochastic modelling}\label{sec.model}

It is generally anticipated that the \TDE\ presented in the previous section for the case of a single pulse can be generalized to a statistically stationary process with a super-position of pulses. The time lags $\tau_x$ and $\tau_y$ are then typically taken as the correlation times that maximizes the cross-correlation functions for measurements at the positions $P_0$, $P_x$ and $P_y$. In this section it is demonstrated that \Eqref{eq.vhat} indeed gives exact results in the case of a super-position of uncorrelated Gaussian pulses that all have the same sizes and velocity components. The derivation provided here will be schematic, with a more comprehensive study following up in a future publication.

Consider a stochastic process given by super-position of $K$ uncorrelated pulses,
\begin{equation}\label{eq.2d}
    \Phi_K(x,y,t) = \sum_{k=1}^{K(T,L)} a_k \varphi\left( \frac{x-v(t-t_k)}{\ell_x}, \frac{(y-y_k)-w(t-t_k)}{\ell_y}\right) ,
\end{equation}
where each pulse with amplitude $a_k$ arrives at $x=0$ and $y=y_k$ at time $t=t_k$, and where $y_k$ and $t_k$ are uniformly distributed random variables. The process has duration $T$ and the vertical domain size is $L$. The pulses move with velocity components $(v,w)$ and have horizontal and vertical sizes $\ell_x$ and $\ell_y$, respectively. All pulses are taken to have the same functional form $\varphi$, which is here assumed to be Gaussian,
\begin{equation}
    \varphi(\theta_x, \theta_y) = \frac{1}{2\pi}\,\exp\left(-\frac{\theta_x^2 + \theta_y^2}{2} \right) .
\end{equation}
This process is temporally stationary and spatially homogeneous. A straight forward average over all random variables gives the mean value of the process,
\begin{equation}\label{eq.Phiave}
    \Phiave = \aave\frac{\ell_x}{v\tauw}\frac{\ell_y}{L} ,
\end{equation}
where the angular brackets $\ave{\cdot}$ represent an average over all random variables, $\aave$ is the average pulse amplitude and $\tauw=T/\Kave$ is the average waiting time between pulses. The factor $\ell_x/v\tauw$ is the ratio of the radial transit time $\ell_x/v$ to the average waiting time and therefore determines the degree of temporal pulse overlap. Similarly, $\ell_y/L$ determines the degree of spatial pulse overlap in the vertical direction. It can furthermore be shown that the variance of the process is given by
\begin{equation}\label{eq.Phirms2}
    \Phi_\text{rms}^2 = \frac{\langle{a^2\rangle}}{2\pi}\frac{\ell_x}{v\tauw}\frac{\ell_y}{L} ,
\end{equation}
showing that the relative fluctuation level $\Phirms/\Phiave$ becomes large when there is little overlap of pulses \cite{garcia_stochastic_2012,garcia_stochastic_2016,garcia_auto-correlation_2017,losada_stochastic_2023}.

The cross-correlation function of this process is defined as
\begin{equation}
    R_\Phi(\xlag,\ylag,\tlag) = \ave{\Phi(x,y,t) \Phi(x+\xlag,y+\ylag,t+\tlag)} ,
\end{equation}
where $\xlag$, $\ylag$ and $\tlag$ are lags in space and time. By performing this average, a lengthy but straight forward calculation gives the cross-correlation function for the process,
\begin{multline}\label{eq.ccf}
    R_\Phi(\xlag,\ylag,\tlag) = \ave{\Phi}^2 \\
    + \Phirms^2 \exp \left( -\frac{1}{2}\left( \frac{\xlag - v\tlag}{\ell_x}\right)^2 -\frac{1}{2}\left( \frac{\ylag - w\tlag}{\ell_y}\right)^2 \right) .
\end{multline}
The time lag $\tlag$ maximizing $R_\Phi$ in \Eqref{eq.ccf} for fixed spatial lags $\xlag$ and $\ylag$ is given by
\begin{equation}\label{eq.tau}
        \tau_{\max{R_\Phi}} = \frac{v \xlag / \ell_x^2 + w \ylag / \ell_y^2}{(v/\ell_x)^2 + (w/\ell_y)^2} .
\end{equation}
It can be demonstrated that if the pulses are tilted such as to be symmetric along the direction of motion, the size dependence drops out and the time of maximum cross-correlation simplifies to
\begin{equation}
    \tau_{\max{R_\Phi}} = \frac{v\xlag+w\ylag}{v^2+w^2} .
\end{equation}
If the process is recorded at three measurement points as illustrated in \Figref{fig.wave_front}, the time lags ${\tau}_x$ and ${\tau}_y$ maximizing the cross-correlations functions at the fixed spatial lags of the given setup can be estimated. In particular, for $\ylag=0$ the cross-correlation function is maximum for time lag ${\tau}_x=v\xlag/(v^2+w^2)$ while for $\xlag=0$ the cross-correlation function is maximum for time lag ${\tau}_y=w\ylag/(v^2+w^2)$.
This is identical to the results in \Eqref{eq.vhat} obtained by geometrical considerations, thus leading to the same expressions for the velocity components.

The results from this stochastic modelling cannot be overemphasized. Firstly, it shows that the three-point calculation of pulse velocities obtained from simple geometric considerations are actually exact in the case of a super-position of uncorrelated pulses which all have the same size and velocity. Secondly, this estimate is independent of the distribution of pulse amplitudes as well as the degree of pulse overlap or the density of pulses. Finally, it confirms that the cross-correlation 
function can indeed be used for \TDE.

\section{Numerical simulations}\label{sec.num}

In order to confirm the theoretical predictions above, we performed numerical realizations of the stochastic process described in the previous section. The Gaussian pulses are taken to be symmetric, $\ell_x=\ell_y=\ell$, and the total speed $u=(v^2+w^2)^{1/2}$ is fixed while the ratio of the velocity components $v/w$ varies between realizations. The vertical domain size is set to $L=10\ell$ and periodic vertical boundary conditions are implemented. The amplitude is for simplicity taken to be the same for all pulses and the total duration of the process is $uT/\ell=10^3$ with $K=10^3$ pulses in total. The resulting fluctuations are measured at a fixed reference point $P_0$, and two auxiliary points $P_x$ and $P_y$ separated by a pulse size $\ell$ in the $x$-direction and in the $y$-direction, respectively. The sampling time is $10^{-2}\ell/u$, thus representing typical experimental conditions with high temporal sampling rate but relatively coarse-grained spatial resolution of the measurements, as detailed in the following section.

\begin{figure}[t]
\centering
\includegraphics[width=7cm]{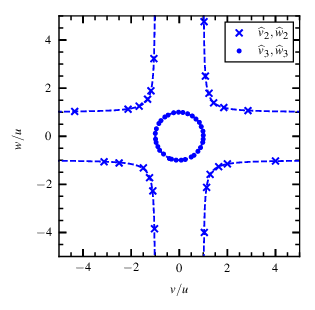}
\caption{Estimated horizontal $v$ and vertical $w$ velocity components from a super-position of uncorrelated Gaussian pulses using the two-point (crosses) and three-point (circles) methods and \TDE\ from cross-correlation functions. Numerous independent realizations of the process are performed for different ratios of the horizontal and vertical velocities at fixed total velocity $u$. The unit circle is the exact velocity while the dashed lines show the erroneous two-point estimate from \Eqref{vw3.hat}.}
\label{fig.estimation_curves}
\end{figure}

Simulations are carried out for various combinations of input velocities $v$ and $w$. For each case, the estimated velocities $\wh{v}$ and $\wh{w}$ are computed using either the two- or the three-point method as described by \Eqsref{eq.vw2.hat} and \eqref{eq.vhat}, respectively. In both cases, the time lags are obtained from the cross-correlation functions. The results of the velocity estimation are presented in \Figref{fig.estimation_curves}, where the filled circles are from the three-point method and the crosses from the two-point method.

As suggested by the stochastic modelling above, the three-point method give correct estimates for both velocity components in all cases, despite a likely marginal spatial resolution for the measurement points which are separated by a pulse size $\ell$ in the numerical simulations. However, the two-point method fails miserably as expected from the discussion in the previous section. When the pulse velocity vector is nearly parallel to one of the coordinate axes, the error for the opposite velocity component can be arbitrarily large. Indeed, the two-point estimate of the velocity vector can according to \Eqref{eq.tau2} be written as $(\wh{v}_2/u,\wh{w}_2/u)=(u/v,u/w)$, which is represented by the dashed lines in \Figref{fig.estimation_curves}. This is an excellent description of the simulation results and demonstrates the failure of the two-point method when one of the velocity components is much smaller than the other.

In order to quantify this, consider the case where we want to ensure the relative error from the two-point estimate of the vertical velocity component $\wh{w}_2$ remains below a value $p$. According to \Eqref{eq.tauy}, this means that $\wh{w}_2/w=1+v^2/w^2<1+p$. It follows that the ratio of the horizontal and vertical velocity components are given by $v^2/w^2<p$, that is, when the error on the estimate of the vertical component is small, $p\ll1$, the horizontal velocity component is significantly smaller than the vertical component. From \Eqref{eq.taux} it furthermore follows that the relative error for the horizontal velocity component is $\wh{v}/v=1+w^2/v^2=1+1/p$. So for a $10\,$\%\ relative error on the vertical component, $p=10^{-1}$, the error for the horizontal velocity component is an order of magnitude larger, $\wh{v}_2/v=11$ when using the two-point estimates. Of course, in practical applications the true velocity components are unknown and there is no way to infer the relative error from the two-point method.

%In order to quantify this, consider the case where we want to ensure the error from the two-point estimate $\wh{w}_2$ remains below $10\%$ of the true value $w$ for the vertical component. According to \Eqref{eq.tauy}, $\wh{w}_2/w=1+v^2/w^2<1.1$. It follows that the horizontal velocity component must be significantly lower, approximately $v/w < 0.3$. However, this is impractical as the real horizontal velocity component $v$ is not known in advance. Moreover, at $v/w=0.3$ the motion is predominantly vertical $w/u=0.95$, but applying the two-point method to the horizontal component would yield $\wh{v}_2/u=(u/v) \approx 3.2$, which is about a tenfold of its actual value.

A comprehensive numerical simulation study has been performed, testing the three-point method for velocity estimation for various pulse functions and distributions of pulse amplitudes, sizes and velocities as well as sampling rates and density of pulses. It is generally found to give accurate estimates of the average velocity components while the two-point method consistently fails to reliably estimate both components. The details of this study will be presented in a separate report.

\section{Experimental measurements}\label{sec.exp}

\begin{figure}[tb]
\centering
\includegraphics[width=6cm]{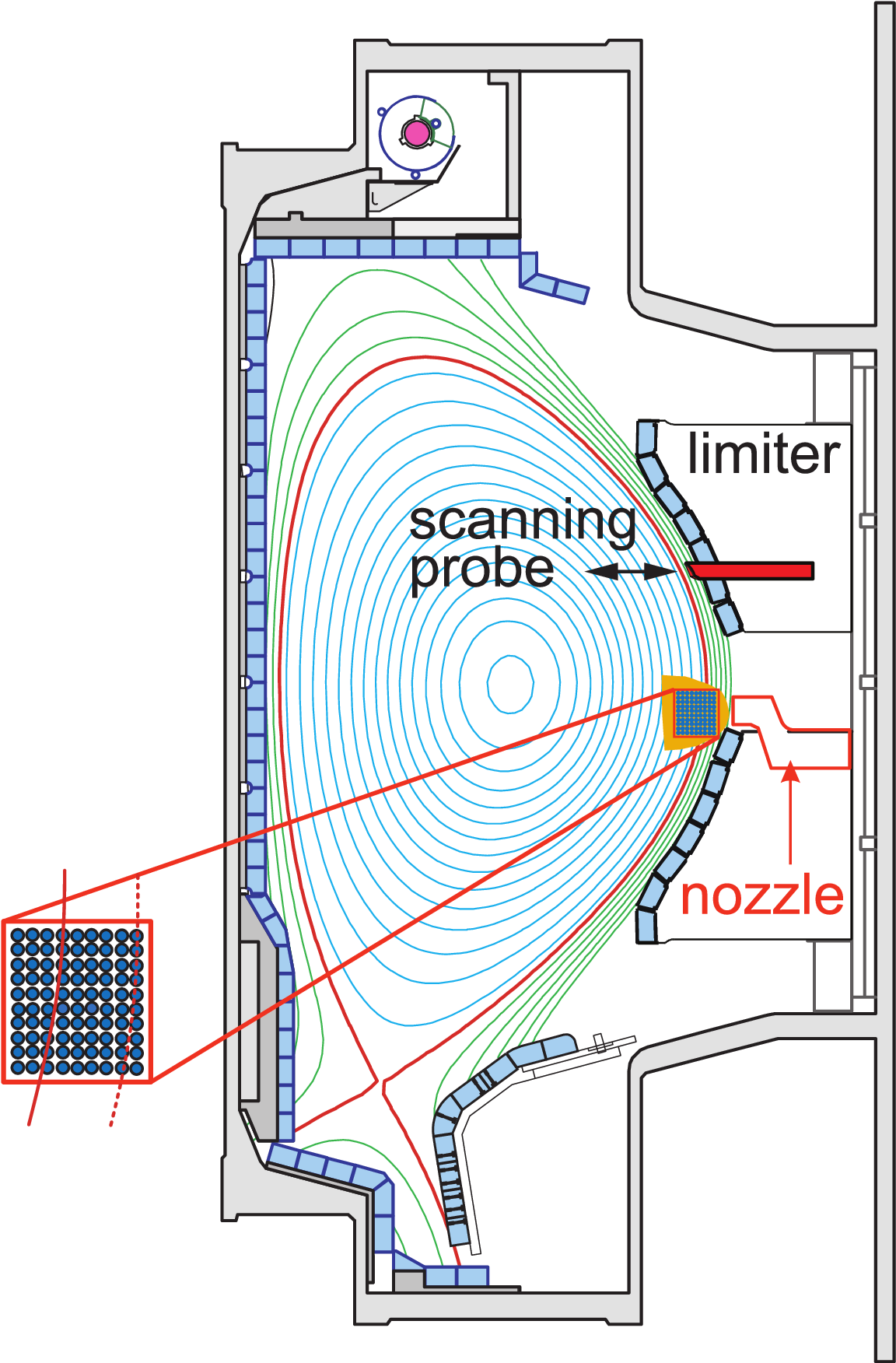}
\caption{Poloidal cross-section of the Alcator C-Mod tokamak showing the closed magnetic flux surfaces in the center of the device (closed cyan curves), the magnetic separatrix and X-point at the lower part of the machine (red curve) and field lines intersecting material surfaces in the scrape-off layer (green curves). The boxes along the plasma boundary show the actual tiles on the divertor and main chamber wall as well as the limiter structures that protect wave antenna and diagnostic systems. Also shown is the location of the gas puff imaging diagnostic system comprised by the $9\times10$ avalanche photo diode view channels whose intensity signals are amplified by neutral gas puff from a nearby nozzle.}
\label{fig:cmod}
\end{figure}

Alcator C-Mod is a compact, high-field tokamak with major radius $R_0=0.68\,\text{m}$ and minor radius $r_0=0.21\,\text{m}$ \cite{hutchinson_first_1994,greenwald_overview_2013,greenwald_20_2014}. The device is equipped with a gas puff imaging diagnostic which consists of a $9\times10$ array of in-vessel optical fibres with toroidally viewing, horizontal lines of sight, as shown in \Figref{fig:cmod} \cite{cziegler_experimental_2010,zweben_invited_2017}. The plasma emission collected in the views is filtered for He I line emission ($587\,\text{nm}$) that is locally enhanced in the object plane by an extended He gas puff from a nearby nozzle, as presented in \Figref{fig:cmod}. Because the helium neutral density changes relatively slowly in space and time, rapid fluctuations in He I emission are caused by local electron density and temperature fluctuations. The GPI intensity signals are therefore taken as a proxy for the plasma density \cite{stotler_neutral_2003,zweben_invited_2017}.

The optical fibres are coupled to high sensitivity avalanche photo diodes and the signals are digitized at a rate of $2\times10^6$ frames per second. The viewing area covers the major radius from $88.00$ to $91.08\,\text{cm}$ and vertical coordinate from $-4.51$ to $-1.08\,\text{cm}$ with an in-focus spot size of $3.8\,\text{mm}$ for each of the 90 individual channels. The measurements presented here were done during the last operational year of Alcator C-Mod. Numerous diodes were broken, leading to lack of data for several channels in the $9x10$ array of measurement points.

The experiment analyzed here was a deuterium fuelled plasma in a lower single null divertor configuration with only Ohmic heating. We present analysis of the plasma fluctuations recorded by the gas puff imaging system for discharge number 1160616016 in the time frame from $1.15$ to $1.45\,\text{s}$. This discharge had a plasma current of $I_\text{p}=0.55\,\text{MA}$, axial magnetic field of $B_0=5.4\,\text{T}$ and a Greenwald fraction of line-averaged density $\overline{n}_\text{e}/n_\text{G}=0.45$, where the Greenwald density is given by $n_\text{G}=(I_\text{p}/\pi r_0^2)10^{20}\,\text{m}^{-3}$ with $I_\text{p}$ in units of MA and $r_0$ in units of meters \cite{greenwald_density_2002}. Analysis of some of these measurement data has previously been reported in Refs.~\onlinecite{kube_comparison_2020} and \onlinecite{ahmed_strongly_2023}.

\begin{figure}[t]
\centering
\includegraphics[width=8cm]{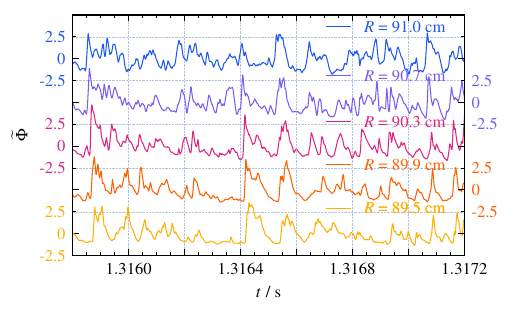}
\caption{Fluctuation time series recorded by the gas puff imaging diagnostic on Alcator C-Mod at different radial positions, reveling radial motion of large-amplitude bursts. Each time series is normalized such as to have vanishing mean and unit standard deviation.}
\label{fig.rawdata}
\end{figure}

Cross-field particle transport in the far scrape-off layer of magnetically confined plasmas are generally attributed to localized blob-like filaments moving radially outwards toward the main chamber wall, resulting in intermittent and large-amplitude fluctuations in the plasma parameters \cite{kube_comparison_2020,ahmed_strongly_2023,garcia_intermittent_2013,theodorsen_relationship_2017,garcia_intermittent_2018,theodorsen_universality_2018}.
%
% Zweben, Maqueda, Terry, Grulke, Myra, Russell, 
%
Such radial motion is evident from the raw measurement data time series at different radial positions, presented in \Figref{fig.rawdata}. Here we present the measured line intensity at $Z=-2.2\,\text{cm}$, where each time series $\Phi(t)$ is normalized such as to have vanishing mean value and unit standard deviation, $\wt{\Phi}=(\Phi-\Phiave)/\Phirms$. Here the mean value $\Phiave$ and the standard deviation $\Phirms^2$ are calculated by a running mean over approximately $1\,\text{ms}$ in order to remove low-frequency trends in the time series. In \Figref{fig.rawdata} there are several large-amplitude events that propagate radially outwards with velocities of the order of $1000\,\text{m/s}$.

\begin{figure*}[tb]
\centering
\includegraphics[width=7cm]{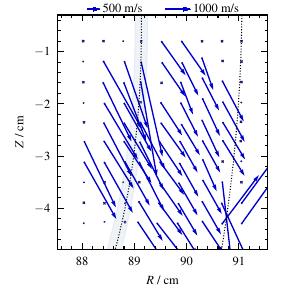}
\hspace*{1cm}
\includegraphics[width=7cm]{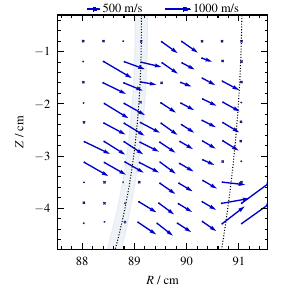}
\caption{Time delay velocity estimation using the cross-correlation function for the two-point method to the left and the three-point method to the right for gas puff imaging measurements on the Alcator C-Mod tokamak.}
\label{fig.tde}
\end{figure*}

The velocity of these fluctuations have been estimated with both the two- and the three-point methods described above, using the lag times that maximizes the cross-correlation function for radial and vertical displacements. For each diode view position, the velocity components are estimated based on correlations with all nearest neighbor diode view positions and averaged radially and vertically. The estimated velocities with these two methods are presented in \Figref{fig.tde}. The crosses correspond to the location of broken diode view positions. No velocity is assigned if either the the cross-correlation function is not unimodal or the time of maximum cross-correlation is less than the sampling time 
$\tlag=5\times10^{-7}\,\text{s}$.

The gray shaded vertical region in \Figref{fig.tde} between major radius of $88.5$ and $89.2\,\text{cm}$ is the location of the last closed magnetic flux surface according to magnetic equilibrium reconstruction. This vertical line separates the confined plasma column to the left and the scrape-off layer to the right of this region. The dotted line at approximately $91\,\text{cm}$ is the location of limiter structures mapped to the gas puff imaging view position.

The two-point method suggests predominantly vertical velocity components through the edge and scrape-off layer regions. However, as anticipated from the theoretical considerations and numerical simulation studies presented above, the three-point method revels that the motion is mainly radial, with velocities up to $1000\,\text{m/s}$. As expected from the above considerations, the two-point method reverses the dominant velocity components and is obviously wrong, as also indicated by the raw data in \Figref{fig.rawdata} which reveals radial motion of the pulses. The results from the three-point method presented here qualitatively agree with previous analysis of Phantom camera data, which has much higher spatial resolution and therefore allows more sophisticated correlation analysis and time delay estimation methods \cite{agostini_edge_2011}. % OTHERS - ZWEBEN, TERRY, ...

\section{Discussion and conclusions}\label{sec.dc}

In this study, we have proposed and evaluated a new method for estimating the velocity components of fluctuation structures that move in two spatial directions.  The method utilizes cross-correlation functions obtained from recordings of a scalar intensity field at three non-aligned measurement points, self-consistently taking into account the two-dimensional nature of the problem. This is demonstrated to give an exact expression for the fluctuation velocity components in a stochastic model given by a super-position of uncorrelated pulses. The accuracy of the method is further demonstrated by numerical realizations of the process, supporting its usefulness also in the case of a random distributions of pulse parameters.

The results from the new three-point method presented here is compared to a standard two-point method. The latter essentially describes the problem as one-dimensional and only give an accurate estimate of the velocity when the two measurement points are perfectly aligned with the direction of motion of the pulses. If the pulses move at a slight angle to any coordinate axis, the two-point method will give a severe overestimate of the corresponding velocity component. Moreover, the two-point method always overestimates both velocity components. This tendency for overestimation has been previously observed in turbulence simulations where the velocity field is known \cite{yu_examination_2007}. Even in that case, only the velocity component in the direction of the two measurement points is correctly estimated, while the perpendicular component would artificially be estimated as infinity. Commonly used Fourier methods with wave number-frequency spectra for one spatial dimension suffer from the same shortcomings as the two-point method \cite{cziegler_experimental_2010}. % OTHERS

Applying the two velocity estimation methods to experimental measurement data from the avalanche photo diode gas puff imaging diagnostic on Alcator C-Mod reveals a striking difference. The two-point method suggests a dominantly vertical motion of the fluctuation structures, which the three-point method demonstrates that the motion is mainly radial as can also be inferred from the raw data time series. The two-point method typically give erroneous results, overestimating the horizontal and vertical velocity components and reversing their ratio. The improved three-point method is therefore likely to give results that are consistent with other analysis methods and diagnostics. In future work this improved velocity estimate method will be used to investigate how fluctuations in the boundary region change with experimental control parameters, improved confinement modes, the role of auxiliary heating, and differences between the far scrape-off layer dominated by motion of blob-like filaments and the closed field line region where vertical wave dynamics is expected to prevail.

In conclusion, our study presents a valuable approach to estimate velocity components in imaging data or stochastic processes characterized by super-posed pulses. By estimating two temporal cross-correlation functions in perpendicular directions, we offer a reliable technique that surpasses the constraints of the standard two-point methods extensively used in previous works. As a final note, it is remarked that the same methodology may be applied with conditional averaging instead of cross-correlation estimation for the time delays. This will be particularly useful for investigating flows associated with large-amplitude fluctuations.

\section*{Acknowledgements}

This work was supported by the UiT Aurora Centre Program, UiT The Arctic University of Norway (2020). Discussions with S.~Ballanger and S.~G.~Baek are gratefully acknowledged. O.~E.~G.\ and A.~D.~H.\ acknowledge the generous hospitality of the MIT Plasma Science and Fusion Center where parts of this work was conducted.

\bibliographystyle{apsrev4-1}
\bibliography{SOL}

\end{document}